\documentclass[10pt]{article}

\usepackage{amsmath,amssymb}
\usepackage{epsfig,lscape,rotating}
\usepackage{graphicx}
\usepackage{multicol}
\usepackage{enumerate}
\usepackage{color}
\usepackage{topcapt}
\usepackage{cite}

\newcommand{\bea}{\begin{eqnarray}}
\newcommand{\eea}{\end{eqnarray}}
\newcommand{\ignore}[1]{}

\newcommand{\bspace}{\!\!\!\!}
\def\met{\mbox{$E{\bspace}/_{T}$}}



\begin{document}

\begin{titlepage}

\vskip.5cm
\begin{center}
{\huge \bf Two Simple W$'$ Models for the Early LHC} \\
\vskip0.4cm
{\huge \bf } 
\vskip.2cm
\end{center}
\vskip1cm

\begin{center}
{\bf Martin Schmaltz and Christian Spethmann} \\
\end{center}
\vskip 8pt

\begin{center}
	{\it 
	Physics Department, Boston University, Boston MA 02215 } \\

\vspace*{0.3cm}

{\tt  schmaltz@bu.edu, cspeth@bu.edu}
\end{center}

\vglue 0.3truecm

\begin{abstract}
\vskip 3pt \noindent
$W'$ gauge bosons are good candidates for early LHC discovery. We define two reference models, one containing a $W'_R$ and one containing a $W'_L$, which may serve as ``simplified models'' for presenting experimental results of $W'$ searches at the LHC. We present the Tevatron bounds on each model and compute the constraints from precision electroweak observables. We find that indirect low-energy constraints on the $W'_L$ are quite strong. However, for a $W_R'$ coupling to right-handed fermions there exists a sizeable region in parameter space beyond the bounds from the Tevatron and low-energy precision measurements where even 50 pb$^{-1}$ of integrated LHC luminosity are sufficient to discover the $W_R'$. The most promising final states are two leptons and two jets, or one lepton recoiling against a ``neutrino jet''. A neutrino jet is a collimated object consisting of a hard lepton and two jets arising from the decay of a highly boosted massive neutrino.
\end{abstract}

\end{titlepage}


\section{Introduction}

$W'$ gauge bosons are attractive candidates for early LHC discovery. A
$W'$ with a TeV scale mass can easily be produced as an s-channel
resonance from quark anti-quark initial states at the LHC.
Generically, $W'$ bosons have significant branching fractions to
leptons leading to an easily observable final state. To understand the
phenomenology of $W'$ bosons it is useful to distinguish two cases:
left-handed $W_L'$ bosons couple to left-handed
quarks and leptons whereas right-handed $W_R'$ bosons
couple to the right-handed fermions.

The goal of our paper is to define two simple $W'$ models as
representatives for each type and to give a broad-brush overview of
their phenomenology. The models are sufficiently simple that much of
the phenomenology only depends on two parameters: the mass of the $W'$
and its universal coupling to quarks and leptons $g_{W'}$. This
simplicity coupled with the possibility to vary both masses and
couplings allows the models to serve as ``simplified models'' for LHC
$W'$ searches in the spirit of \cite{simplified}. The idea is that
bounds on similar models \cite{Mohapatra:1974gc,Mohapatra:1974hk,Mohapatra:1980yp,Chivukula:2006cg,Barger:1980ix,Barger:1980ti,
Georgi:1989ic,Georgi:1989xz,Malkawi:1996fs,Li:1981nk,He:2002ha,LH_original,Littlest_Higgs,LHreviews} 
can be obtained with relative
ease by theorists by converting the bounds on the corresponding
simplified models as determined by the experiments.

The simple models presented here are also sufficiently attractive to
be taken seriously as specific new physics models. We determine the indirect constraints 
\cite{delAguila:2010mx, Hsieh:2010zr, Rizzo:2006nw}
from precision electroweak measurements and LEP2 on
$M_{W'}-g_{W'}$ parameter space of each model and compare them to the
direct bounds from the Tevatron \cite{d0:2007bs,CDFjj,Abazov:2008vj} and the direct
LHC reach \cite{atlas,cms, Ferrari:2000sp,Gninenko:2006br}.
\footnote{For a classic paper on $W'$ phenomenology see 
\cite{Keung:1983uu}, more recent analyses for the LHC are
\cite{WpOther}.}
The left panel in Figure 1 shows that the 
indirect constraints on $W'_{L}$ models are much stronger than the Tevatron
bounds for intermediate and large values of the gauge coupling
$g_{W'}\gtrsim 0.5$. For example, for $g_{W'_L}$ equal to the
Standard Model $SU(2)$ coupling the LHC requires an integrated luminosity of order
10 fb$^{-1}$ to probe beyond the indirect bounds. For smaller
couplings, $g_{W'} \lesssim 0.45$, about 1 fb$^{-1}$ of data is
sufficient to discover a $W'_L$ with a mass up to 1.7 TeV.
In contrast, $W'_R$ bosons with couplings  $g_{W'} \simeq 0.5$ are
only loosely constrained by precision measurements and may be as light
as the Tevatron bound $\gtrsim 1$ TeV. As a result, the early LHC will
probe well beyond existing bounds. For example, a $W'_R$ with mass
just beyond the Tevatron limit can already be discovered with the 2010 
LHC data set corresponding to $\approx$ 50 pb$^{-1}$ of integrated luminosity per experiment,
and would therefore be a ``supermodel'' in the sense of \cite{Bauer:2009cc}.

\begin{figure}
\centering
 \includegraphics[width=0.45\textwidth]{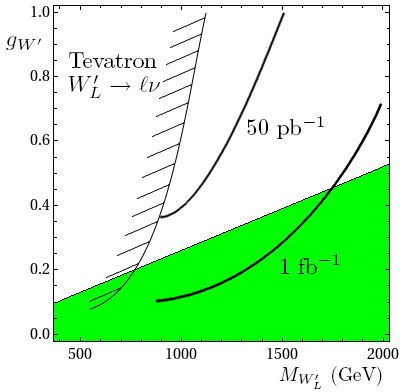}
 \includegraphics[width=0.45\textwidth]{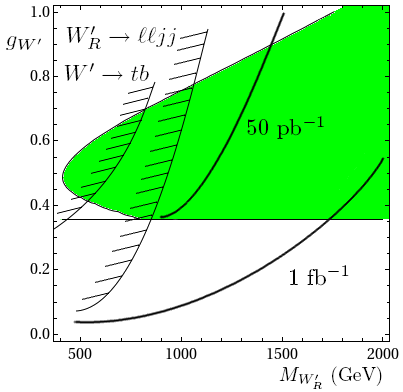}
 \caption{Experimental constraints and LHC reach for the left-handed
(left) and right-handed (right) \mbox{$W$ prime} as functions of the
simplified model parameters $M_{W'}$ and $g_{W'}$.
The plots show limits from direct searches at the Tevatron (hashed
contours), the region favored by electroweak precision
fits at \mbox{95 \% C.L.} (green/gray region), and the LHC reach at
\mbox{$\sqrt{s}=7$ TeV}
for \mbox{50 pb$^{-1}$} and \mbox{1 fb$^{-1}$}
of integrated luminosity.}
\label{fig:introplots}
\end{figure}

The $W'_L$ model is based on the extended electroweak gauge group
$SU(2)_1 \times SU(2)_2 \times U(1)_Y$ breaking to $SU(2)_L \times
U(1)_Y$ at the TeV scale. The Standard Model fermions and Higgs
transform under $SU(2)_1$ at high energies. $W'_L$ is a linear
combination of the two charged gauge bosons in $SU(2)_1 \times
SU(2)_2$, it couples to all left-handed fermions and the Higgs doublet
with equal strength. The constraints from low energy measurements on the $W'_L$ 
mass are quite strong because amplitudes mediated
by the $W'_{L}$ interfere with amplitudes mediated by the Standard
Model $W$. Such interference effects are only suppressed by
$(M_W/M_{W'})^2$ and give the bounds shown in the left panel of Figure
1. At hadron colliders one searches for the $W'_L$ with the signature
of a high $p_T$ lepton and missing energy from the decay $W'_L
\rightarrow l \bar \nu$. The Tevatron can only compete with
precision bounds at very small couplings $g_{W'} \lesssim 0.2$ whereas
the LHC with 1 fb$^{-1}$ surpasses precision bounds also for more
strongly coupled $W'_L$ bosons up to $g_{W'} \lesssim 0.45$.

The $W'_R$ model is based on the extended electroweak gauge group
$SU(2)_L \times SU(2)_R \times U(1)_{B-L}$ broken to $SU(2)_L \times
U(1)_Y$ at the TeV scale. The $W'_R$ boson couples only to
right-handed quarks and leptons and does not interfere with the
Standard Model $W$. As a result, the $W'_R$ contributions to
low-energy measurements are suppressed by   $(M_W/M_{W'})^4$ and give
much weaker constraints. Constraints from the $Z'_R$ which
accompanies the $W'_R$ are still significant. We see in the right
panel of Figure 1 that precision measurements allow $W'_R$ masses as
small as 500 GeV for $g_{W'} \simeq 0.5$. The lower bound on the mass
depends on the $Z'_R$ couplings which in turn depend on
the embedding of the Standard Model Higgs boson
in the model. This embedding introduces an additional parameter which has been
projected out in Figure 1.

At hadron colliders, $W'_R$ resonances can be discovered in hadronic
final states from  $W'_R \rightarrow jj$, $W'_R \rightarrow t \bar b$ 
or in leptonic channels via $W'_R \rightarrow l \bar N \rightarrow l \bar ljj$. 
The hadronic channels are generally more difficult but become
important when the leptonic channel is closed by phase space because
the right-handed neutrino $N$ is heavier than the $W'_R$. Assuming
that the leptonic channel is open there are two important kinematic
regimes. The right-handed neutrino $N$ may be very heavy with a mass
of several 100 GeV so that all final state objects are highly
energetic and well separated. In this case one would look for the
final state $ l \bar l jj$ (or the lepton number violating $lljj$ final state if the right-handed neutrino
has a Majorana mass). Alternatively, the right-handed neutrino
might be as light as the lower bound on its mass from the
non-observation of heavy neutrinos in $Z$ decays, $M_{N}\gtrsim 45$
GeV. In this case the right-handed neutrino and its decay products
are boosted and merge into a single massive ``neutrino jet'' ($j_N$). This is a
new object consisting of two collimated subjets and an embedded hard lepton.
Relatively light right-handed neutrinos with
masses in the range 45 - 200 GeV thus require a search for final states with a 
very energetic charged lepton recoiling against the neutrino jet.
Here the muonic channel is probably the
most promising because it is easy to identify the embedded hard lepton
inside the neutrino jet.

In Section 2 of this paper we define the two ``simplified models''
with $W'_L$ and $W'_R$ gauge bosons. Section 3 is dedicated to the
calculation of contributions to precision electroweak observables and
plots of the allowed parameter spaces after fitting to precision data.
For the case of the $W'_R$ the precision fit depends on the embedding
of the Standard Model Higgs into the model and we discuss the
different possibilities in some detail. Readers who are mainly
interested in our models as simplified models are invited to skip
the Section on precision electroweak constraints because such
constraints can be avoided with sufficient model building
ingenuity. In Section 4 we show the existing Tevatron bounds.
Published $W'$ bounds from the Tevatron refer to specific  $W'$ models
with fixed values of the coupling constant $g_{W'}$. We ``reprocess''
the searches to extend their validity to our more general simplified
models. In Section 5 we present LHC cross sections as a function of
$M_{W'}$ and $g_{W'}$ for different interesting final states.
Particularly promising are the $l \bar l jj$ and $l j_N$
signatures from decays of $W'_R$ to heavy and light right-handed
neutrinos, respectively.

The Appendices are dedicated to model building issues. In the first
Appendix we consider alternate representations for the Higgses which
break $SU(2)_R$. Using larger $SU(2)$ representations it is possible
to raise the $Z_R'$ mass relative to the $W_R'$ mass. This alleviates
the precision electroweak bounds from $Z_R'$ exchange and allows even
lighter $W_R'$ bosons at the expense of slightly more complicated
Higgs sectors. In the second Appendix we outline different scenarios
for fermion mass generation, paying special attention to naturalness
of neutrino masses.

\section{Description of $W'$ Models}

\subsection{The minimal $W_L'$ Model}

We start our discussion by defining a simplified model that implements a $W_L'$. 
The electroweak gauge group is extended to 
\[ SU(2)_1 \times SU(2)_2 \times U(1)_Y . \]
All Standard Model fermion doublets and the Higgs doublet transform under $SU(2)_1$. 
A bifundamental Higgs field with VEV
\bea \langle \Delta \rangle = \frac1{\sqrt{2}} \; \begin{pmatrix} f & 0 \\ 0 & f \end{pmatrix} , \eea
breaks the two $SU(2)$ factors to the diagonal $SU(2)_L$. Before electroweak symmetry breaking,
the heavy gauge boson eigenstates are degenerate with mass
\bea M_{Z_L'}^2 = M_{W_L'}^2 = \frac14 \; f^2 \; (g_1^2+g_2^2) . \eea
The mixing matrix between the heavy $W_L'$ and the Standard Model $W$ is 
\bea
\begin{pmatrix} W \\ W_L' \end{pmatrix}
=
\begin{pmatrix} \cos \theta_L & - \sin \theta_L \\ \sin \theta_L & \cos \theta_L \end{pmatrix}
\begin{pmatrix} W_1 \\ W_2 \end{pmatrix},
\eea
where
\bea \sin \theta_L = \frac{g_1}{\sqrt{g_1^2+g_2^2}}, \qquad \cos \theta_L = \frac{g_2}{\sqrt{g_1^2+g_2^2}} . \eea
The Standard Model $SU(2)$ gauge coupling is the combination
\bea \frac{1}{g^2} = \frac{1}{g_1^2} + \frac{1}{g_2^2}, \eea
which implies $g_1,g_2 \ge g$. The heavy gauge bosons couple to the Standard Model doublets with  
universal coupling strength
\bea g_{W'} = g \tan \theta_L . \eea
In the special case $g_1=g_2$ the $W_L'$ is therefore a sequential $W'$, with coupling strength 
equal to the Standard Model gauge coupling $g$.
Below the $W_L'$ and $Z_L'$ mass scale, the fermion and scalar field content of the minimal 
$W_L'$ model is identical to the Standard Model.

\subsection{The minimal $W_R'$ Model}

The simplest extension of the SM with a right-handed $W'_R$ has the extended electroweak gauge group
\[ SU(2)_L \times SU(2)_R \times U(1)_X , \]
with fermion charges under $U(1)_X$ equal to $(B-L)/2$.
To keep our analysis general, we do not assume left-right symmetry, thus the couplings $g_L$ and $g_R$ are independent. Hypercharge is the combination
\bea Y = T^3_R + X \eea
with coupling strength
\begin{equation}
\label{eq:gprime}
\frac1{g'^2}=\frac1{g_R^2}+\frac1{g_X^2}.
\end{equation}
In particular, this equation implies $g_X,g_R > g'$.
We assume the minimally required scalar sector at the $SU(2)_R$ breaking scale, which consists of 
a single $SU(2)_R$ doublet. 
This scalar doublet carries 
a $U(1)_X$ charge of +1/2, so that one of the components has zero hypercharge. 
The scalar acquires a VEV 
\bea \langle H_R \rangle = \frac1{\sqrt{2}} \; \begin{pmatrix} 0 \\ f \end{pmatrix}  \eea
which breaks $SU(2)_R \times U(1)_X$ down to $U(1)_Y$. 
Neglecting electroweak symmetry breaking, the heavy gauge boson masses are
\begin{equation}
\label{eq:mwpmzp} 
M^2_{W_R'} = \frac14 \, f^2 g_R^2, \qquad
M^2_{Z_R'} = \frac14 \, f^2 \, (g_R^2+g_X^2).
\end{equation}
in complete analogy to the Standard model. The neutral gauge boson mixing matrix is
\bea 
\begin{pmatrix} Z_R' \\ B \end{pmatrix}
=
\begin{pmatrix} \cos \theta_R & - \sin \theta_R \\ \sin \theta_R & \cos \theta_R \end{pmatrix}
\begin{pmatrix} W^3_R \\ X \end{pmatrix},
\eea
where $X$ is the $U(1)_X$ gauge boson and 
\bea \sin \theta_R = \frac{g_X}{\sqrt{g_X^2+g_R^2}}, \qquad \cos \theta_R = \frac{g_R}{\sqrt{g_X^2+g_R^2}}. \eea
The coupling strength $g_R$ of the $W_R'$ boson is bounded from below by the hypercharge coupling $g'$, 
whereas the $W_L'$ coupling can be arbitrarily small. The required fermion representations in the $W_R'$ model 
(including a right-handed neutrino and a gauge singlet $N$) are shown in Table \ref{tab:fermions}.

\begin{table}[t!]
\begin{center}
\topcaption{Fermion representations in the minimal $W_R'$ model. $Q^c$ and $L^c$
are left-handed spinors that transform as $\bar{2}$'s under $SU(2)_R$. 
$N$ is a gauge singlet fermion which was included to allow a Dirac mass 
with the right-handed neutrino $\nu^c$.}
\small
\begin{tabular}{c|cc|cc|c}
\rule[-1ex]{0ex}{3ex} & $X$ & $T^3_R$ & $Y$ \\ \hline
$Q$ \rule[-2.5ex]{0ex}{6ex} & 1/6 & 0 & 1/6 \\ \hline
$Q^c=\begin{pmatrix} u^c \\ d^c \end{pmatrix}$ & -1/6 & $\mp 1/2$ \rule[-2.5ex]{0ex}{6ex} &
  $\begin{pmatrix} -2/3 \\ 1/3 \end{pmatrix}$ \\ \hline
$L$ \rule[-2.5ex]{0ex}{6ex} & -1/2 & 0 & -1/2 \\ \hline
$L^c=\begin{pmatrix} \nu^c \\ e^c \end{pmatrix}$ & 1/2 & $\mp 1/2$ \rule[-2.5ex]{0ex}{6ex} & 
  $\begin{pmatrix} 0 \\ +1 \end{pmatrix}$ \\ \hline
$N$ \rule{0ex}{3.5ex} & 0 & 0 & 0 
\end{tabular}
\normalsize
\label{tab:fermions}
\end{center}
\end{table}

\subsection{Higgs Sector in the $W_R'$ Model}

To keep the analysis general, we include two Higgs electroweak symmetry breaking representations,
\begin{enumerate}
\item An $SU(2)_L \times SU(2)_R$ bidoublet $\phi$ with zero $U(1)$ charge, and
\item An $SU(2)_L$ left-handed doublet $H_L$ with $U(1)$ charge +1/2.
\end{enumerate}
After $SU(2)_R$ breaking, the bidoublet decomposes into two $SU(2)_L$ doublets with hypercharges $\pm 1/2$. 
The two fields $\phi$ and $H_L$ acquire VEVs
\bea \langle \phi \rangle = \frac{1}{\sqrt{2}} \; \begin{pmatrix} \kappa & 0 \\ 0 & \kappa' \end{pmatrix},
\qquad \langle H_L \rangle = \frac{1}{\sqrt{2}} \; \begin{pmatrix} 0 \\ v_L \end{pmatrix} .
\eea
Since both $\langle \phi \rangle$ and $\langle H_L \rangle$ contribute to electroweak 
gauge boson masses, the VEVs have to satisfy the relation 
$\kappa^2 + \kappa'^2 + v_L^2 = v^2 = (246 \mbox{ GeV})^2$.
Important special cases are $\kappa=\kappa'=0$ or $v_L=0$ for which the precision 
electroweak constraints are the same as in a model which contains only
one of the electroweak symmetry breaking fields ($H_L$ or $\phi$).

Fermion masses may be generated from renormalizable Yukawa interactions with
the bifundamental Higgs doublet $\phi$ and/or from dimension five operators with the $H_L$ and $H_R$ doublets,
\bea \mathcal{L}_{Y} = \lambda_{\phi q} Q \phi Q^c
+ \frac{\lambda_{hq}}{\Lambda} \; Q H_L \tilde{H_R} Q^c , \eea 
plus possibly couplings of these fields with epsilon contractions of $SU(2)_L$ indices.
Additional Yukawa interactions may be present to generate first and second generation quark masses, as well 
as lepton masses. However, the fit to precision data as well as the $W_R'$ phenomenology is independent of quark 
and charged lepton masses. 
Right-handed neutrinos have to be massive and decay promptly to a lepton plus two jets.
Their masses can be Dirac or Majorana, affecting the $W_R'$ decay signature, but not the precision fit. 
For completeness, we briefly discuss fermion masses in the Appendix.

\section{Electroweak Precision Constraints}

\subsection{Constraints on the $W_L$ model}

Following Han and Skiba \cite{Han:2004az}, we quantify the impact of new 
physics on electroweak precision observables by calculating the 
coefficients of 21 dimension six operators of Standard Model 
fermions and the Higgs field. 
In the $W_L'$ and $W_R'$ models the constrained operators are induced by 
the exchange of scalars and of the heavy gauge bosons $W'$ and $Z'$. 

Let us turn our attention to scalar exchange first. Because
of Lorentz invariance, scalar fields can couple to
fermion bilinears, but not to covariant derivatives of the Standard Model Higgs. 
Therefore scalar exchange can only generate the four-fermion
operators considered by Han and Skiba. Of those, only
the ones involving first generation quarks and leptons are strongly restricted 
by electroweak precision measurements. In our models, the contribution of scalar 
exchange to such operators is small because the Yukawa coupling 
coefficients are proportional to the fermion masses.
We will therefore concern ourselves only with contributions 
due to heavy gauge boson exchange.  

In the $W_L$ model before electroweak symmetry breaking, 
the three heavy gauge bosons ${W_L'}^{\pm}$ and $Z_L'$ transform as a triplet under $SU(2)_L$. 
They couple to the current
\bea J_{W_L}^{\mu,a} = g_{W'} \left[ \ell^\dagger \bar{\sigma}^\mu T^a \ell
+ q^\dagger \bar{\sigma}^\mu T^a q 
+ ( i h^\dagger T^a D^\mu h + \mbox{h.c.} ) \right] , \eea
where $T^a=\frac12 \tau^a$ are the $SU(2)_L$ group generators.
The effective Lagrangian is then
\bea \mathcal{L}_{\mathrm{eff}} = \mathcal{L}_{\mathrm{SM}} - \frac12 \; \frac1{M_{W_L'}^2} J_{\mu}^a J^{\mu,a} , \eea  
so that the coefficients of the triplet operators in Han and Skiba's formalism are universal
\bea a_t = - \frac14 \frac{g_{W_L'}^2}{M_{W_L'}^2} . \eea

The results of the electroweak fit of the $W_L'$ model are shown as
shaded contours in Figure \ref{fig:LHCpt}.

\subsection{$Z'$ Currents and Charges in the $W_R'$ model}

In the $W_R'$ model the situation is more complicated, since the $W_R'$ and $Z_R'$ 
do not transform as an irreducible representation of the Standard Model gauge group.
It is therefore necessary to
examine $Z_R'$ and $W_R'$ exchange separately. As we will see below, only the $Z_R'$ 
produces strongly constrained operators. We start by calculating the
couplings of the fermions and the two Higgs representations to the $Z_R'$.

The fermion contribution to the $Z_R'$ current is
\begin{equation}
\label{eq:fermion_charges} 
J_{Z_R'}^\mu(f) = \sum_f  \left( g_R \cos \theta_R T^3_R 
- g_X \sin \theta_R X \right) f^\dagger \bar{\sigma}^\mu f . 
\end{equation}

The operator description by Han and Skiba is formulated in terms of a single Higgs doublet, 
so at first glance it would appear that additional scalar representations
cannot be accommodated in this formalism.
However, the VEV is the only component of the Higgs field that has been
measured so far and is therefore relevant for the electroweak fit. 
The left-handed doublet contribution to the $Z_R'$ current is
\bea
J_{Z_R'}^\mu (H_L) =  - i \frac12 g_X \sin \theta_R \; H_L^\dagger D^\mu H_L + \mbox{h.c.} 
\begin{matrix} \\ \Rightarrow \\ \mbox{\tiny EWSB} \end{matrix} 
- \frac14 v_L^2 (g'^2+g_L^2)^{1/2} g_X \sin \theta_R Z^\mu . 
\eea
Similarly, the contribution of the Higgs bidoublet is
\bea
J_{Z_R'}^\mu (\phi) = i g_R \cos \theta_R \; \mbox{tr} \left[ (T_3^R \phi)^\dagger (D^\mu \phi) \right] 
+ \mbox{h.c.}
\begin{matrix} \\ \Rightarrow \\ \mbox{\tiny EWSB} \end{matrix} 
\; \frac14 ({\kappa}^2+{\kappa'}^2) \, (g'^2+g_L)^{1/2} g_R \cos \theta_R Z^\mu .
\eea
We can therefore represent the scalar sector contribution to the 
$Z_R'$ current in terms of a single Higgs field $h$ with VEV $v$ 
as in the Standard Model,
\begin{eqnarray} 
J_{Z_R'}^\mu(h) & = & \frac12 \left( \frac{\kappa^2+\kappa'^2}{v^2} \; g_R \cos \theta_R - \frac{v_L^2}{v^2} \; 
g_X \sin \theta_R \right) \; \left[ i (h^\dagger D^\mu h) + \mathrm{h.c.} \right] \nonumber \\
& = & \frac12 \; \sqrt{g_R^2+g_X^2} \; \left( \cos^2 \theta_R - \frac{v_L^2}{v^2} \right) 
\; \left[ i (h^\dagger D^\mu h) + \mathrm{h.c.} \right] 
\end{eqnarray}

\begin{figure}
 \centering
  \includegraphics[width=0.26\textwidth]{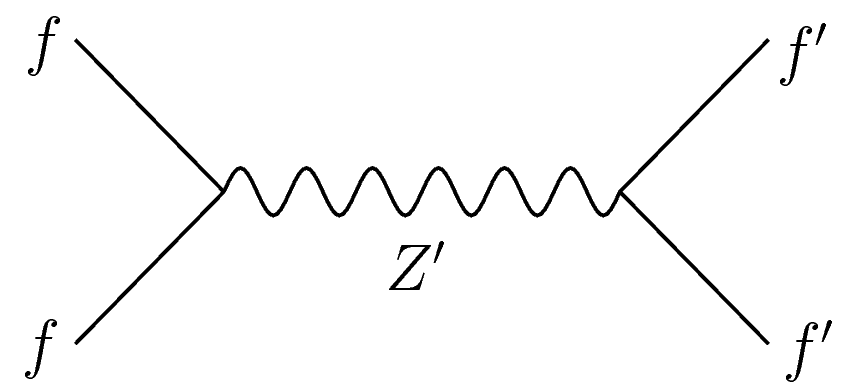}
  \hfill
  \includegraphics[width=0.3\textwidth]{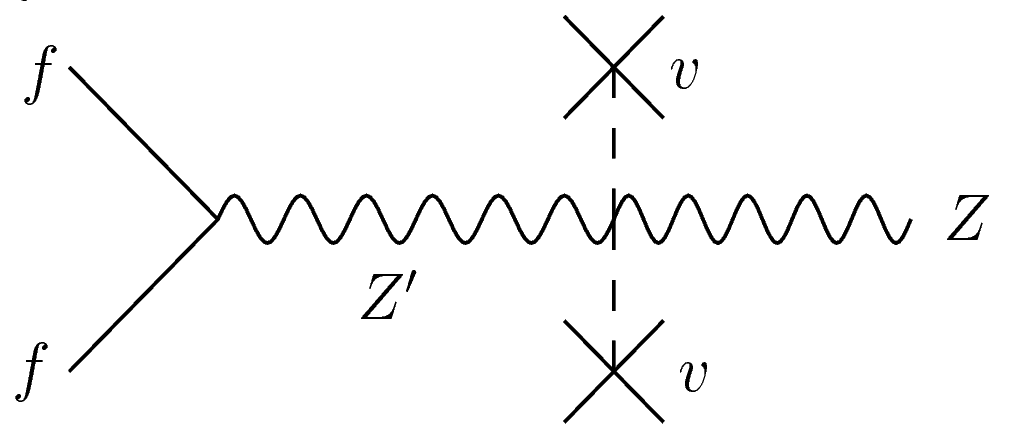}
  \hfill
  \includegraphics[width=0.35\textwidth]{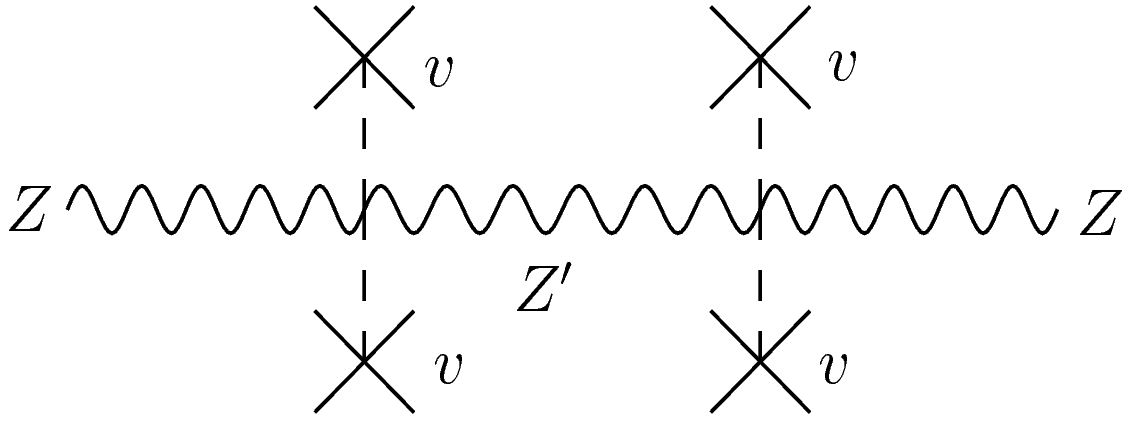} 
  \caption{Dimension six operators from $Z'$ exchange: a) Four fermion couplings, b) Corrections
to the fermion-Z coupling, c) Mass shift for the Z, which contributes to the T-parameter.}
\label{fig:zpdiagrams}
\end{figure}

$Z_R'$ exchange generates three different kinds of dimension six operators, as shown in Fig. \ref{fig:zpdiagrams}. 
The first of those are four fermion interactions, which are represented by the operators
\bea
a_{ij}^{(s)} \mathcal{O}_{ij}^{(s)} 
= - \frac4{f^2} \; \left( \rule{0ex}{3ex} \cos^2 \theta_R T^3_{R,i} - 
\sin^2 \theta_R X_i \right) \; 
\left( \rule{0ex}{3ex} \cos^2 \theta_R T^3_{R,j} - 
\sin^2 \theta_R X_j \right) \; f_i^\dagger \bar{\sigma}^\mu f_i \; f_j^\dagger \bar{\sigma}_\mu f_j .
\eea
The operators that modify fermion-Z couplings are 
\begin{equation}
\label{eq:hfoperators}
a_{hi}^{(s)} \mathcal{O}_{hi}^{(s)} 
= - i \; \frac2{f^2} \; 
\left( \cos^2 \theta_R - \frac{v_L^2}{v^2} \right) \;
\; \left( \rule{0ex}{3ex} 
\cos^2 \theta_R T^3_{R,i} - \sin^2 \theta_R X_i \right) \;
(h^\dagger D_\mu h) \; f_i^\dagger \bar{\sigma}^\mu f_i + \mbox{h.c.} . 
\end{equation}
Finally, the $Z$ mass is shifted by the T operator 
\bea
a_h \mathcal{O}_h = - \frac2{f^2} \; 
\left(\cos^2 \theta_R - \frac{v_L^2}{v^2} \right)^2  
|h^\dagger D_\mu h|^2.
\eea

Exchange of the $W_R'$ does not generate four fermion operators with 
strong constraints from electroweak precision measurements. However,
one might be concerned that mixing with the Standard model $W$ shifts the mass of the 
$W$ mass eigenstate and therefore contributes to the $\rho$ parameter.
Diagonalizing the mass-squared matrix for the charged gauge bosons
and expanding in powers of $v^2/f^2$, we find
\bea \frac{\delta M_W}{M_W} = - \frac{2 \kappa \kappa'}{f^2} + \mathcal{O}(v^4/f^4) . \eea
To reproduce the large top/bottom mass hierarchy, at least one of
the two bidoublet VEVs must be much smaller than the electroweak
symmetry breaking scale $v$. 
Thus the $W_R'$ mass shift contribution to the $\rho$ parameter
has an additional suppression beyond the $v^2/f^2$ present in the
contribution to $\delta \rho$ from $Z_R'-Z$ mixing.
We will therefore neglect $W_R'$ exchange.

The dimension 6 operators generated by $Z_R'$ exchange depend on 3 model parameters: the scale of $SU(2)_R$ breaking $f$, the right-handed gauge boson mixing angle $\tan \theta_R = g_X/g_R$, and the fraction of the electroweak symmetry breaking vev squared $v_L^2/v^2$ contained in $H_L$. 
Since we will be interested in $W'$ phenomenology, it is convenient to change variables from $f$ and $\theta_R$ to the more experimentally relevant $g_R=g'/\sin \theta_R$ and the $W_R'$ mass $m_{W_R'}=g_R f/2$. For the remaining parameter, $v_L^2/v^2$, we focus on three representative cases $v_L=0, v_L=v$, and $v_L=v/\sqrt{2}$.

\begin{figure}
\centering
  \includegraphics[width=0.49\textwidth]{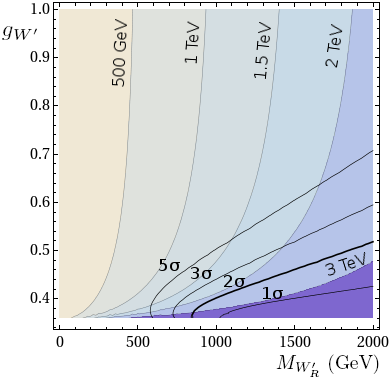}
  \hfill
  \includegraphics[width=0.49\textwidth]{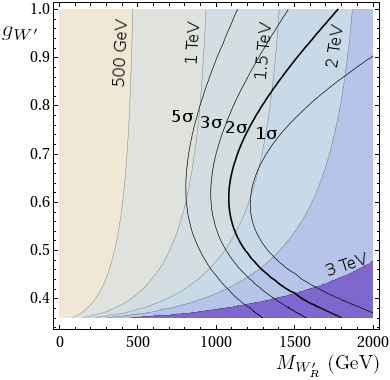}
  \caption{Contour plots of electroweak precision exclusion levels and the 
$Z_R'$ mass in the $W_R'$ model as a function of the $W_R'$ mass 
and the right-handed coupling constant $g_{W'}=g_R$. Left: Purely bidoublet VEV ($v_L=0$), Right:
purely left-handed doublet VEV ($v_L=v$).}
\label{fig:ewp}
\end{figure}

\subsection{Bidoublet Higgs}

Let us first consider the case where only the bidoublet has a VEV, $v_L=0$. 
Since we can neglect scalar
exchange contributions in the electroweak fit, this limit is equivalent
to removing $H_L$ completely.
The result of our fit to the precision observables for this case is shown in the left plot in Fig. \ref{fig:ewp}.
Note that there is a minimum theoretically allowed value for the $W_R'$ coupling $g_R=g'$. This value 
corresponds to the limit $g_X \rightarrow \infty$. The Figure shows
that for a given $M_{W_R'}$, the smallest possible values of $g_R$ (and very large $g_X$) are preferred.
This is easy to understand by noticing that the strongly constrained operators 
$\mathcal{O}_{hf}$ and  $\mathcal{O}_{h}$ in Eq. \ref{eq:hfoperators}
are both proportional to powers of $\cos \theta_R = g'/g_X$
In the preferred region of parameter space the $Z_R'$ is very heavy and  
out of reach of the early LHC.

In the rest of this subsection we will briefly elaborate on the precision fit in the limit of large $g_X$ and examine whether a large value for the $g_X$ coupling constant is reasonable from 
a model building perspective.

In the limit of $g_X \gg g'$, the electroweak constraints are particularly 
easy to understand because only the four fermion operators in Eq. \ref{eq:hfoperators} 
survive. Furthermore, we have in this limit
\bea a_{\mathrm{ff}'} \propto \frac{g_X^2}{M_{Z_R'}^2} = \frac{4g_X^2}{(g_X^2+g_R^2) f^2} 
\to \frac{4}{f^2} . \eea
The most stringent constraint on these four-fermion operators comes from measurements of $e^+e^- \to e^+e^-$
scattering at LEP II \cite{LEP} and bounds the operator 
\bea \mathcal{O}_{4e} = \frac{2 \pi}{(\Lambda_{ee}^-)^2} 
\bar{e} \gamma_\mu e \, \bar{e} \gamma^\mu e . \eea
This operator is precisely the one generated by exchange of the $Z_R'$  in our model in the limit
of large $g_X$ because the $Z_R'$ couplings become left-right symmetric.
The experimental bound on $\Lambda_{ee}^-$ implies a limit of $f \geq 5.1$ TeV at 
the 95 \% confidence level.
Since $g_R \approx g'$ we then obtain the minimum allowed $W_R'$ mass as
$M_{W_R'} \geq fg'/2 \approx$ 900 GeV, which reproduces the lower bound on
$W_R'$ masses in our plot. 

We now briefly examine what a reasonable upper bound on the allowed values for $g_X$ might be.
Large values for the gauge coupling of a $U(1)$ gauge group are a concern
as they increase in the UV due to running and require new physics before they become
non-perturbatively large. One possibility for the this new physics is that the $U(1)$ gauge group is embedded in
a non-abelian group, thereby turning around the sign of the beta function. Here, a particularly nice
possibility for the embedding of $U(1)_X$ would be to consider near-TeV scale gauge ``unification"
\bea SU(3)_{\mathrm{color}} \times U(1)_X \subset SU(4) , \eea
motivated by the fermion charges which naturally fit into this group.  
Using the $SU(4)$ normalization of $g_X = \sqrt{3/2} \; g_3$ and neglecting 
the running of the gauge couplings between the unification scale and the TeV scale, we obtain $g_X \approx 1.5$. 
This value of the coupling lies in the preferred region of the model parameter space.

\subsection{Left-Handed Higgs}

We now consider the opposite case, a Higgs VEV only for the $SU(2)_L$ doublet $H_L$, $v_L=v$. This is equivalent
to eliminating the bidoublet Higgs field from the theory.
\ignore{From a model-building perspective, the remaining setup is appealing 
because the field content is now left-right symmetric, even though the right-handed
symmetry is broken at a higher scale than the left-handed symmetry, $f \gg v$. }
The $T$-operator from $Z'$ exchange becomes
\bea
a_h \mathcal{O}_h = - \frac2{f^2} \sin^4 \theta_R |h^\dagger D_\mu h|^2,
\eea
and the fermion-Higgs coupling operators are
\bea
a_{hi}^{(s)} \mathcal{O}_{hi}^{(s)} 
= - i \; \frac2{f^2} \; \sin^2 \theta_R \; \left( \rule{0ex}{3ex} 
\cos^2 \theta_R T^3_{R,i} - \sin^2 \theta_R X_j \right) \;
(h^\dagger D_\mu h) \; f_i^\dagger \bar{\sigma}^\mu f_i + \mbox{h.c.} . 
\eea
Which region of parameter space is now preferred? The Higgs couplings 
are suppressed for small $\sin \theta_R = g'/g_R$, thus preferring small values of $g_X$. 
However, the $Z'$ couplings to fermions in Eq. \ref{eq:fermion_charges}  
prefer a value of $\tan \theta_R$ which minimizes
the couplings to charged leptons because they
are most strongly constrained by LEP measurements. This competition leads to a broader
distribution of the preferred values of $g_R$.
The resulting exclusion plot is shown in the
right half of Fig. \ref{fig:ewp}.
The electroweak fit predicts a more strongly coupled $W_R'$,  
$M_{Z'}$ relatively close to $M_{W'}$, and small $Z'$ couplings to charged leptons.

\subsection{Hybrid Higgs}

\begin{figure}
\centering
  \includegraphics[width=0.49\textwidth]{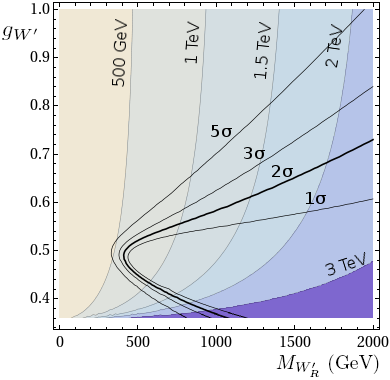}
  \includegraphics[width=0.49\textwidth]{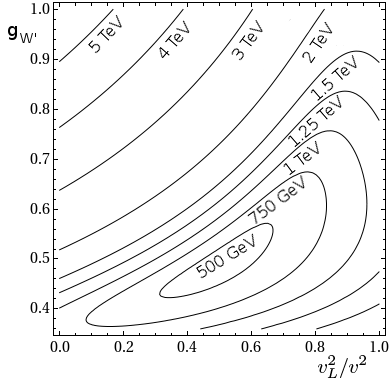}
  \caption{Left: Electroweak constraints on the Hybrid Higgs model for $(v_L/v)^2=1/2$.
  Right: Minimally allowed $W'$ mass (in units of GeV) as a function of $(v_L/v)^2$ and $g_R$.}
\label{fig:hybrid}
\end{figure}

Finally, we discuss the general case where both electroweak breaking VEVs are non-zero.
In the left half of Fig. \ref{fig:hybrid} we set the Higgs VEVs for 
both fields $\phi$ and $H_L$ to $v/\sqrt{2}$ and plot electroweak precision
constraints on the parameter space of the resulting model. 
There is a narrow region where a $W_R'$ with a mass $\approx$ 400 GeV is allowed. 
This region of parameter space appears fine-tuned. To investigate the degree of fine-tuning necessary
to obtain such a small $W'$ masses, we plot the minimal allowed $W_R'$ mass as function
of the model parameters $(v_L/v)^2$ and $g_R$
(right half of \mbox{Fig. \ref{fig:hybrid}}).

From the figure it is evident that the precision fit prefers 
a narrow band in parameter space centered around $v_L \approx v_\phi\approx v/\sqrt{2}$ and $g_R \approx g_X \approx \sqrt{2} g'$.
In this region, the $Z'$ couplings to the 
two Higgs representations cancel, and there are consequently no constraints from the
$T$ parameter or from shifts of the $Z$ coupling to fermions. In addition, setting the gauge coupling
constants equal decouples the right-handed electron from the $Z'$, since for this
field the $X$ and $T^3_R$ charges are identical. Electroweak precision measurements
therefore place weak bounds on such a $Z'$, allowing the scale $f$ and the $W_R'$ mass to be low.

In this model there is no symmetry reason for why the VEVs of scalars in
different representations should be equal or why the coupling constants of the $SU(2)_R$ and
$U(1)_X$ gauge groups should be equal. Thus maybe $W_R'$ masses of order 1 TeV corresponding
to more generic regions of parameter space should be considered more natural.
However, it is remarkable that a $Z'$ boson with the relatively strong
couplings $g_R=g_X=\sqrt{2} g'$ is consistent with the precision data for masses as low as $m_{Z_R'}\sim 570$ GeV, allowing $M_{W_R'}\sim 400$ GeV.

As we will see in the next section, direct searches
from the Tevatron imply much stronger limits on the $W_R'$ mass. 
The fit to precision electroweak observables of our model with 
$(v_L/v)^2=1/2$, $g_R=g_X$ and $M_{W'_R}=1$ TeV is 
better than the Standard Model fit, with $\Delta \chi^2=-2.0$.

\section{Tevatron Exclusion Limits}

\subsection{Tevatron Limits on $W'_L \to l \nu$}

The CDF and D\O~experiments at the Tevatron have searched for signatures of $W'$ bosons
decaying in both leptonic and hadronic channels. We will first compare the 
results from the most recent D\O~search for $W_L' \to e \nu_e$ \cite{d0:2007bs} with 
electroweak precision limits from LEP2. In the Tevatron study, a set of inclusive 
single electron and missing energy cuts were applied on a data \
sample of $\approx$ 1 fb$^{-1}$. After reconstruction, two events with transverse mass 
$m_T > 500$ GeV were found. Taking into account statistical and systematic backgrounds, 
the D\O~collaboration concluded that a sequential $W_L'$ with SM strength coupling to electrons and
Standard Model neutrinos can be excluded for $M_{W_R'}$ up to 1 TeV at the 95 \% C.L.

To understand which other values of $W_L'$ masses and couplings this search was sensitive to,
we used {\sc madgraph} \cite{MadGraph} to 
simulate 1000 parton level events at each of 210 model points with $W_L'$ masses between 500 GeV and 1500 GeV and 
gauge coupling constants between 0.05 and 1. Since initial state radiation 
is not included in {\sc madgraph}, the $W_L'$ resonance is always 
produced without transverse momentum in the simulated events. 
The transverse momenta of the lepton and neutrino are then 
equal and opposite, and the transverse mass of the $W_L'$ is
\bea m_T(W_L') = 2\, {p_T(l)}. \eea
Counting events with lepton $p_T > 250$ GeV ($m_T > 500$ GeV) and $|\eta_e|<1.1$, 
we find that with 1 fb$^{-1}$ of integrated luminosity 3.0 events would be expected for a sequential 
$W_L'$ at $M_{W_L'}=$ 1 TeV. Approximating the signal acceptance as independent of phase space, 
we use this number of events to find the 95 \% C.L. exclusion limit for different coupling 
strengths and masses. 
The resulting Tevatron exclusion limit is shown as a hashed contour in Figure \ref{fig:introplots}. 
In the most common models (e.g. KK gauge bosons modes, Little Higgs) a $W_L'$ is accompanied by a $Z'$ with equal 
mass and coupling strength, and LEP2 data excludes such a $Z_L'$ with a mass up to 
2.6 TeV. Our derived Tevatron bound is only competitive
for coupling constants less than $g_R=0.2$. At such small coupling, the early LHC cannot reach beyond the Tevatron bound.

\begin{figure}
\centering
  \begin{minipage}{0.35\textwidth}
  \includegraphics[width=\textwidth]{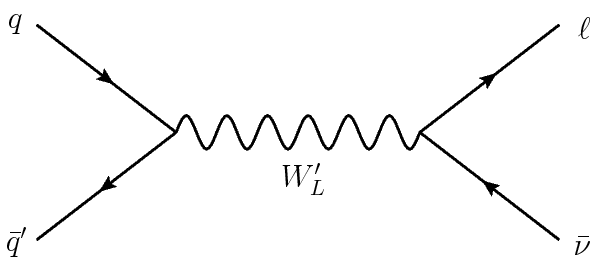}
  \end{minipage}
  \hfill
  \begin{minipage}{0.55\textwidth}
  \includegraphics[width=\textwidth]{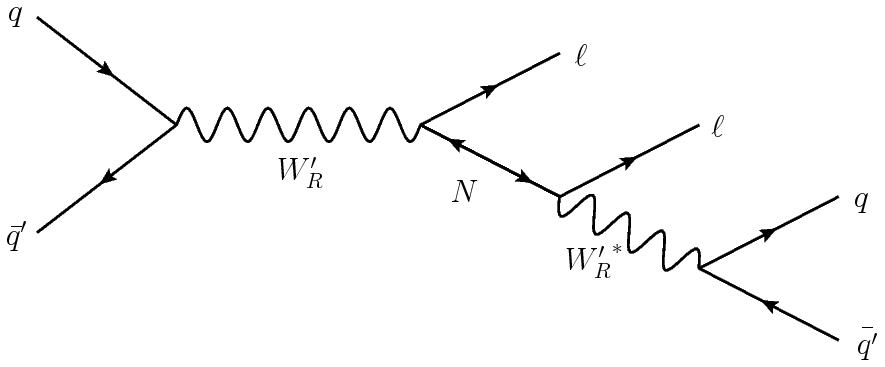}
  \end{minipage}
  \caption{Left: Decay of the $W_L'$ to a lepton and anti-neutrino.
  Right: Decay of the $W_R'$ to a same sign dilepton and two jets via a right-handed 
  Majorana neutrino and a virtual $W_R'$.}
\label{fig:decays}
\end{figure}

\subsection{Tevatron Limits on $W'_R \to ee jj$}

The results from the above search do not apply to the $W_R'$ decaying to a 
lepton and an unstable heavy neutrino, since such events do not produce missing energy.
The Tevatron collaborations have not published any search results for $W_R' \to ll jj$
decays. There are however several searches for leptoquark pair production, which shares the 
same final $eejj$ and $\mu \mu jj$ states \cite{D0LQ1,D0LQ2}. 
Because of the different event topology, those results are not directly applicable to  
the $W_R'$ simplified model. However, the production and decay of $W_R'$ resonances (with
subsequent decay of heavy right-handed neutrinos) would be noticeable in those data analyses 
as a surplus of events with large dilepton invariant mass and large scalar transverse momentum sum.
We therefore re-interpreted the Tevatron leptoquark search to extract limits on the $W_R$ 
resonance mass and coupling strength. 

As will be described in more detail in the LHC section, the $\mu \mu jj$ final state is more sensitive
to the $W_R'$ decay signal because of reduced lepton isolation requirements. It would therefore
appear that using the D\O~second generation leptoquark search would yield stronger limits on the
simplified model parameter space. 
However, the $\mu \mu jj$ study uses a multivariate analysis to 
distinguish the leptoquark signal from background
including cuts on $\mu j$ invariant masses, which are not meaningful for right-handed neutrino decay.
The first generation leptoquark search instead uses cuts on the dilepton mass 
and the total transverse momentum. We therefore apply our signal to the selection 
cuts of the first generation leptoquark search.

To estimate the maximum mass of a $W_R'$ that would have been visible at the Tevatron, we perform 
a scan over $W_R'$ masses in the range 500-1200 GeV and right-handed neutrino masses up to $M_{W_R'}$ 
using the left-right model included in {\sc pythia} 8.1\cite{Pythia}. 
The analysis was performed at the parton level. Detector effects are approximated by smearing the
energies of electrons and jets according to a Gaussian with standard deviation
\bea \sigma_E = a \cdot \sqrt{E} \oplus b \cdot E , \eea
where the Energy is measured in GeV and $\oplus$ stands for adding in quadrature.
For the constants we choose
\bea a_{\mathrm{jet}}=0.80, \quad b_{\mathrm{jet}}=0.05, \quad a_{e} = 0.20, \quad b_e=0.01 , \eea
following \cite{Drees:1996rw}.
After energy smearing, we apply a set of kinematic and isolation cuts:
\begin{itemize}
\item For each electron-jet pair and jet-jet pair, $\Delta R > 0.5$
\item For any electron and any jet, $p_T > 25$ GeV
\item At least one of the two electrons must be central, $|\eta_e|<1.1$, 
      and both electrons must be within the electromagnetic calorimeter, $|\eta_e|<1.1$ or $1.5<|\eta_e|<2.5$
\item Both jets must be within the hadronic calorimeter, $|\eta_j|<2.5$
\end{itemize}
Finally, we apply the background rejection cuts of 
the first generation leptoquark analysis of $m_{ee}>110$ GeV and $\sum p_T>$ 400 GeV.

\begin{figure}
\centering
\includegraphics[width=0.49\textwidth]{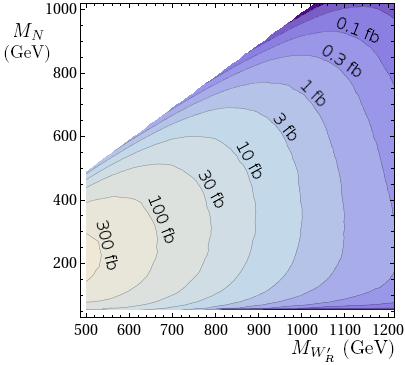}
\hfill
\includegraphics[width=0.49\textwidth]{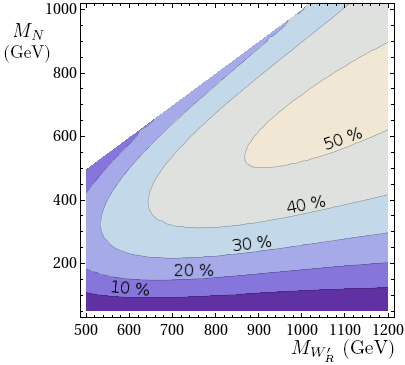}
\caption{Left: Effective Tevatron cross section for $W_R' \to jj ee$ signal events after first generation
leptoquark search cuts as described in the text. The gauge coupling constant is assumed to be $g_R=g$. The
LEP excluded region ($M_N<45$ GeV) has been omitted.
Right: Signal acceptance as a function of the same variables. 
}
\label{fig:leptoquark}
\end{figure}

In the left part of Figure \ref{fig:leptoquark}, we plot the cross 
section for signal events that pass the leptoquark search cuts as a function of the
$W_R'$ and the heavy neutrino mass for the left-right symmetric case $g_L=g_R$.
In the right half of the figure we show the signal acceptance. 

In the Introduction (right half of Fig. \ref{fig:introplots}), we show the resulting exclusion line 
for different values of the right-handed gauge coupling, assuming that the right-handed 
neutrino mass is close to the optimal value of 
300 GeV. In making the introduction plot we assumed that five signal events are
sufficient for the signal to be visible.

\subsection{Tevatron Limits on $W'_{L/R} \to jj/tb$}

Even if the decay of the right-handed $W_R'$ to leptons is kinematically allowed, the 
Tevatron searches would have been unable to see the signal in the following cases:
\begin{enumerate}
\item 
If $m_N \approx M_{W_R'}$, the branching ratio to leptons becomes small, reducing
the effective cross section. 
\item  
For neutrino
masses below 100 GeV, the two jets and the electron from the neutrino decay
can not be sufficiently separated. The leptoquark search is therefore not sensitive
to a $W_R \to eejj$ decay signal in this mass region. 
\end{enumerate}
To assure that a $W_R'$ with suppressed leptonic decay modes has not been missed,
it is necessary to compare the above result with the search for $W'$ resonances in 
the $jj$ and $tb$ final states, which arise from $W'$ decays to quarks. 
The CDF collaboration has searched for resonances
in the dijet invariant mass spectrum, and found that a $W'$ with Standard Model 
couplings can be excluded up to \mbox{$M_{W'}=800$ GeV} \cite{CDFjj}. Since the width
of the resonance depends on the coupling strength, this result cannot be extended 
to other values of $g_R$ in a straightforward way.
The D\O~collaboration has searched for the decay of a resonance into top+bottom quarks,
and found a limit of 740 GeV for a $W'$ with SM coupling $g$ \cite{Abazov:2008vj}. The mass limits from  
this search for different values of the gauge coupling strength are included in 
Fig. \ref{fig:introplots}.

\section{LHC Reach}

For the left-handed $W_L'$ and the right-handed $W_R'$ which we consider in this paper the 
$W'$ couplings to fermions are universal. This makes the determination of branching fractions 
particularly simple. 
Neglecting fermion masses, the branching fractions of the $W'$ are 1/12 for 
each generation of leptons and 1/4 for each generation of quarks. 
The comparatively large branching fraction to leptons of 1/6 ($e N_e$ and $\mu N_\mu$ combined)
makes the $W'$ a favorable search target for the early LHC. 
In the following we will concentrate on the leptonic decay mode
\[ {W_{R/L}'}^{\pm} \to l^\pm N . \]
Here $N$ is either the Standard Model left-handed neutrino in the case of a $W_L'$ 
or the heavy right-handed neutrino in the left-right model. 

The ATLAS collaboration has published a detailed study of both $W_L'\to l \nu$
and $W_R'\to l N \to ll jj$ events at the 14 TeV LHC \cite{atlas}.
For the right-handed $W_R'$ two model points were investigated in detail,
including all Standard Model backgrounds and a full detector simulation.
The gauge coupling constant was assumed to be equal for the left-handed and right-handed
$SU(2)$ groups.

To compare the early LHC reach with the results from our electroweak fit,  
it is necessary to generalize this analysis to arbitrary values of the 
$W_{L/R}'$ mass, right-handed neutrino mass and gauge coupling strength.
We therefore again perform a parton level scan over the parameter space.
The details of our analysis will be described below.

\subsection{$W_L' \to l \nu$ at the LHC}

Let us discuss the decay of a $W_L'$ to a lepton and a massless neutrino first. 
The expected signal of a hard lepton plus missing energy 
must be compared with the most relevant background from
the high $p_T$ tail of the $W$ decay spectrum. We assume that all 
other Standard Model backgrounds such as $t\bar{t}$ and QCD dijet misidentification
can be sufficiently reduced by appropriate cuts on the hadronic event activity,
which is minimal in leptonic $W_L'$ decays.\footnote{It should be noted that 
cosmic ray muons can falsely produce a $\mu$ + MET signal. However, we will not 
consider such events in our analysis, assuming that a good understanding of the 
detector and the use of timing information will reduce those backgrounds sufficiently.}

\begin{figure}
\centering
\begin{minipage}{0.505\textwidth}
\includegraphics[width=\textwidth]{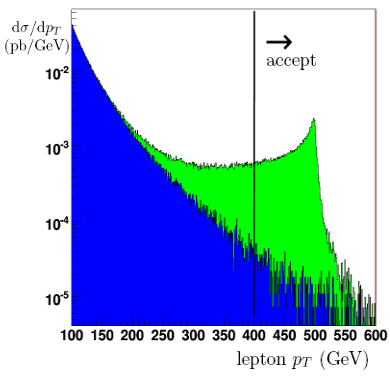}
\end{minipage}
\begin{minipage}{0.485\textwidth}
\includegraphics[width=\textwidth]{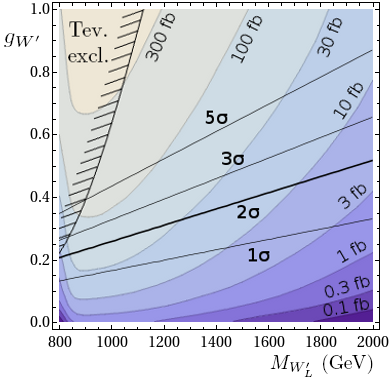}
\end{minipage}
\caption{Left: Lepton $p_T$ distribution for a $W_L'$ signal with $M_{W_L'}=1$ TeV 
and the irreducible Standard Model $W\to e \bar{\nu}_e$ background (blue). Right: 
Cross section $\times$ branching ratio $\times$ acceptance for the $W_L' \to e \nu_e$
signal at the 7 TeV LHC. Also shown are the limits from the Tevatron search and the electroweak
precision fit.}
\label{fig:LHCpt}
\end{figure}

Using {\sc madgraph}, we simulated the production and decay of $W_L'$ bosons with masses between 
800 GeV and 2 TeV and couplings up to $g_{W'}=1$. At each model point, 1000 events 
were generated at the parton level.
To estimate the background from the high $p_T$ tail of $W \to l \nu$ events,
we simulated 270k leptonic $W$ decays with {\sc madgraph}, applying a lepton $p_T$ cut of 
$100$ GeV at the generator level. The cross section of this Standard Model
background falls of steeply with $p_T$, such that only 2.8 background 
events with lepton $p_T > 400$ GeV are expected in each of the electron and
muon channels for 1 fb$^{-1}$ of LHC data. 
The results of our background simulation, compared with the expected signal for a 1 TeV $W_L'$ 
boson with gauge coupling strength $g_{W'} \approx g'$, can be seen in the left half of 
Fig. \ref{fig:LHCpt}.

We then simulated the production and decay of $W_L'$ boson events with masses between 800 GeV and 
2 TeV and couplings up to $g_{W'}=1$ to find the effective cross section 
$\times$ branching fraction $\times$ signal efficiency.
The resulting effective cross sections are shown as shaded contours 
in the right half of Fig. \ref{fig:LHCpt}.

We also investigated the integrated luminosity required for discovery with 50 pb$^{-1}$ and 1 fb$^{-1}$. 
For any integrated luminosity below 300 pb$^{-1}$, less than one background event is 
expected. We therefore assume that in this case 5 signal events after the lepton $p_T$ cut are 
sufficient to establish the presence of new physics. With 1 fb$^{-1}$, the background is 
non-vanishing. We assume purely statistical errors and demand at least $5 \sqrt{N_{BG}}=8.3$ signal events. 

The $W_L'$ has to be weakly coupled to avoid electroweak precision constraints. 
Our results suggest that such a sequential $W_L'$ can not be found with the first 50 pb$^{-1}$ 
of early LHC data. However, with 1 fb$^{-1}$ of integrated luminosity, 
the LHC experiments become sensitive to a $W_L'$ in the allowed region of parameter space 
with $W_L'$ mass up to $\approx$ 1.5 TeV and gauge coupling constant $g_{W'}$ smaller than $g'$.
A $W_L'$ with Standard Model like couplings is excluded by electroweak precision
constraints up to several TeV (see the left panel of Figure 1.).

\subsection{$W_R' \to ll jj$ at the LHC}

In the $W_R'\to l N$ two-body decay, the lepton is produced with fixed momentum
\bea 
p_{l} = \frac1{2 M_{W_R'}} \left( M_{W_R'}^2 - m_{N}^2 \right).
\eea
The lepton $p_T$ distribution therefore exhibits a Jacobian peak at this value, 
corresponding to the events where the lepton is emitted perpendicularly to the beam pipe
in the $W'_R$ rest frame.
For massless quarks and non-zero neutrino mass, the branching ratio for each leptonic decay mode is
\bea 
BR(W_R'\to e N) = \frac{(1-x^2)^2 (2+x^2)}{3(1-x^2)^2(2+x^2)+18},
\eea
where $x=m_N/M_{W_R'}$ is the ratio of the neutrino and $W_R'$ masses.

The heavy neutrino is unstable and decays via a virtual $W_R'$ to another lepton and two jets. For large right-handed neutrino masses the resulting two leptons and two jets will be widely separated and easy to identify. Combining the jets with one of the leptons one obtains the invariant mass of the neutrino, combining both leptons and the jets one reconstructs $M_{W'_R}$.  

For the right-handed $W_R'$ decay to two hard leptons and two jets, 
we assume that all Standard Model 
backgrounds can be eliminated with appropriate cuts. This 
is certainly the case for same sign dilepton events, resulting
from the decay of heavy Majorana neutrinos. For 
oppositely charged leptons plus jets, the most relevant 
backgrounds are $Z/\gamma+$jets, $t\bar{t}$ and
pairs of electroweak gauge bosons. However, the 
the small integrated luminosity of less than 1 fb$^{-1}$ 
at the early LHC implies that cuts on the dilepton mass
and on the scalar transverse momentum sum result in 
an effectively background free sample.
We therefore estimate the early LHC reach by demanding 
at least five signal events after basic kinematic cuts.

\begin{figure}
\centering
\includegraphics[width=0.49\textwidth]{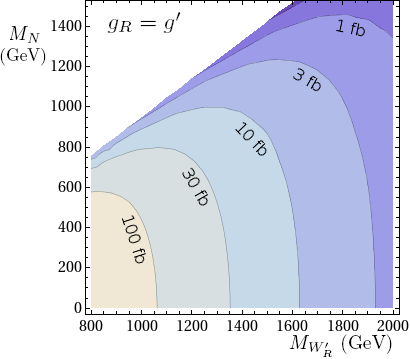}
\includegraphics[width=0.49\textwidth]{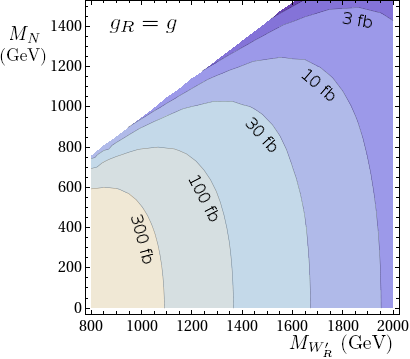}
\caption{Effective cross section $\times$ branching fraction for $W_R'\to eN$, as a function 
of the $W_R'$ mass and the right-handed neutrino mass. Left: For coupling strength $g_{W'}=g'$, Right:
for $g_{W'}=g$.}
\label{fig:eNsignificance}
\end{figure}

\subsection{$W_R' \to l j_N$ at the LHC}

If $M_{W_R'} \gg M_N$, the lepton and two jets from the heavy neutrino decay are emitted
in a cone with opening angle
\bea \Delta R \approx \frac{M_N}{M_{W_R'}} .  \eea
The event reconstruction at ATLAS and CMS by default implements 
the anti-$k_T$ jet algorithm with cone sizes of $\Delta R_j = 0.4-0.7$. 
If the typical separation of the neutrino decay products is of this size or less, 
the two quarks can no longer be identified
as separate objects but will instead appear as a single hadronic jet \cite{Ferrari:2000sp}. 

The ``neutrino jet'' is an object which will be interesting to search for at the LHC. 
It can be easily identified if it contains a hard muon, but even a boosted neutrino 
decaying into an electron and two hadronic subjets 
will produce a distinct signature that can be resolved using jet substructure techniques. Given the relative importance of this region in parameter space and the expected cleanliness of the final state signal (a hard lepton recoiling against a boosted neutrino jet) we believe that this case deserves a designated study. For our plots we have assumed that the $\bar l j_N$ final state is easy to identify and has no Standard Model backgound, and our discovery contours correspond to 5 signal events.

How light can a right-handed neutrino be and have avoided detection so far?
Since the right-handed neutrino is uncharged under the Standard Model gauge group, its 
only interaction with ordinary matter is through $W-W_R'$ and $Z-Z'$ mixing. 
The mixing angle $\alpha$ is of order $\mathcal{O}(v^2/f^2)$, so that the 
decays $Z \to N N$ and $W \to l N$ are suppressed by a factor of 
$\alpha^2 \approx 10^{-4}$ compared to the corresponding Standard Model decay
modes into left-handed neutrinos. 
The only process in which right-handed neutrinos would have been produced in
observable quantities is $Z$ decay at LEP 1.
The experimental bound on $M_N$ is therefore the kinematic limit 
of $M_Z/2 \approx 45$ GeV.

\subsection{$W_R' \to jj/tb$ at the LHC}

It is also important to search for $W_{L/R}'$ bosons via decays to quarks which are necessarily
present with large branching fractions if the $W'$ can be produced at a Hadron collider.
In the case of a decay to light quark jets, the QCD cross section 
for dijet production is orders of magnitude larger than a possible dijet $W'$ signal. Furthermore, 
the shape and overall rate of the large invariant mass tail of the QCD background is not known 
accurately enough to establish the presence of a signal over the Standard Model background
with the limited dataset available after the first two years of LHC running.

A search for $W'$ decays into third generation quarks appears more promising, because the 
SM production of a top/bottom quark pair necessarily requires $W$ exchange and is much smaller. 
The background to this signal is dominated by QCD events such
as mistagged $q\bar{q}$ pair production with W-Strahlung or pair production of top quarks. A study of $W'$ decays
into boosted third generation quarks using jet substructure techniques would
be interesting, but is beyond the scope of this paper.

\section{Conclusions}

$W'$ searches are interesting examples of early physics for the LHC. The absence of strong indirect constraints from low energy measurements particularly motivates searches for the final states $l\bar ljj$ and $l \bar{j}_N$ from $W'_R$ decays, already with 50 fb$^{-1}$ of integrated luminosity.
Note that in the case of a Majorana mass for the right-handed neutrino the final state leptons can have the same or opposite sign charge 
with equal branching fractions.

We stress that searches for all possible $W'$ final states 
\[ 
W_L' \to \left\{ \begin{matrix}l + \met \\ t \bar b \\ jj \end{matrix} \right. 
\qquad \qquad
W_R' \to \left\{ \begin{matrix}l\bar ljj \\ l \bar{j}_N \\ t \bar b \\ jj \end{matrix} \right. 
\]
are interesting with larger data sets (1-10 fb$^{-1}$). This is especially true if one keeps in mind that the final states which we have 
emphasized most may be closed due to phase space. In less universal models than the ones we have considered the $W'$ may not couple 
to leptons at all so that only $jj$ and $tb$ final states are available. Finally, $W'$ bosons may also decay to $WZ$ and $Wh$ with branching fractions which are suppressed by powers of $v/f$ and are therefore less promising.

\section*{Acknowledgements}
We wish to acknowledge useful conversations with Andy Cohen, Liam Fitzpatrick and Brock Tweedie and especially wish to thank 
Jesse Thaler for collaboration during the early stages of this project. This research is supported by the US Department of 
Energy under grant DE-FG02-01ER-40676.

\appendix

\section{Splitting $W_R'$ and $Z_R'$ Bosons with Tensors}

The Standard Model tree level relation 
\bea \frac{M_W}{M_Z} = \cos \theta_W, \eea
i.e.~$\rho=1$, is a consequence of electroweak symmetry breaking with $SU(2)$ doublet Higgses. If the Higgses were taken to transform in higher $SU(2)$ representations, the relation would no longer be true and the $Z$ would be relatively heavier. We can make use of this fact in the $W_R'$ model to increase the 
mass splitting between $Z_R'$ and $W_R'$. This is desirable because the precision electroweak constraints on $W_R'$ models are dominated by $Z_R'$ exchange.

To employ this in the context of the $SU(2)_R \times U(1)_X \to U(1)_Y$ model, at least two 
scalar fields are required. 
First, to raise the $Z'$ mass we add a complex triplet field 
$\Delta$ with VEV
\bea
\langle \Delta \rangle = ( 0, 0, k/\sqrt{2} ) 
\eea
and a $U(1)$ charge of +1, so that hypercharge is unbroken.
In order to write the Yukawa couplings for the Standard Model fermions we also need a doublet scalar $H_R$ with a VEV $(0,f/\sqrt{2})$. The tree level masses of the gauge bosons are then
\bea  M_{Z_R'}^2 = \frac18  \, (g_R^2+g_X^2) \, (4 k^2+f^2)   \eea
and
\bea M_{W_R'}^2 = \frac18  \, g_R^2 \, (2 k^2 + f^2). \eea
In the limit $k \gg f$, the indirect EW precision constraint on the $W_R'$ mass is therefore reduced by a factor $1/\sqrt{2}$. More generally, by using a scalar transforming in the $D$-dimensional representation of $SU(2)_R$ the indirect constraint on the $W'_R$ is reduced by a factor of $1/\sqrt{D-1}$.

\section{Fermion Masses and $SU(2)_R$ Higgs Representations}

In this Appendix we briefly discuss the issue of fermion masses and flavor. In models with full left-right symmetry at the TeV scale it is quite difficult to arrange for quark and lepton masses because the left-right symmetry predicts identical mass ratios for up- and down-type quarks as well as for charged and neutral leptons. Here, we have much more freedom because we are not interested to preserve even an approximate left-right symmetry at the TeV scale. 

For example, we may obtain the Standard Model quark masses from the following operators
\bea
\mathcal{L}_{Y} &=& \lambda_u Q \phi Q^c + \frac{\lambda'}{\Lambda^2} (Q \phi H_R)( H_R^\dagger Q^c ) \\
&\rightarrow&  \lambda_u Q H_u u^c + (\lambda_u +  \frac{\lambda'}{2}\frac{f^2}{\Lambda^2})\,  Q H_d d^c  \ .
\eea
The use of the ``projector" $H_R$ onto the right-handed down quark allows for a splitting between up- and down-type masses without introducing tree level flavor changing neutral currents from Higgs exchange. As usual in theories with $SU(2)_R$ gauge bosons there are potentially disastrous box diagrams with $W_L$ and $W_R$ exchange which generate contributions to neutral meson mixing. The constraints from the neutral Kaon system for generic large right-handed flavor mixing angles are particularly severe and rule out $W'_R$ bosons near the TeV scale. These constraints are avoided if the right-handed CKM matrix takes on a specific form \cite{Langacker:1989xa,Buras:2010pz}. We will ignore the constraint on the $W'_R$ mass and assume that excessive K-Kbar mixing is avoided by fortuitous right-handed mixing.   

Even so, obtaining satisfactory lepton masses is a little tricky. The problem is that the operator
 \bea
\lambda_e L \phi L^c \rightarrow  \lambda_e  L H_d e^c  +  \lambda_e L H_u \nu^c 
\label{eq:e_masses}
\eea
adds unwanted large Dirac masses for neutrinos in addition to the charged lepton masses. Since the neutrino masses are 10 orders of magnitude smaller than the tau lepton mass, even a 4-loop radiatively generated coupling of the form Eq.~(\ref{eq:e_masses}) would give neutrino masses which are too large. \ignore{ Note that we already used the charge-conjugate $\tilde \phi$ in order to suppress neutrino masses relative to charged lepton masses by a factor of $\tan\beta$.}
Using the projector $H_R$ does not help much in this case because the operator with $H_R H_R^\dagger$ radiatively generates the operator without $H_R$. Similarly, attempting to construct a model in which only one of the two Higgs doublets $H_u$ and $H_d$ obtains a VEV is problematic because radiative corrections generate a VEV for the other Higgs.

We circumvent this potential problem by including three generations of gauge singlet neutrinos $N$ to marry off the unwanted right-handed neutrinos $\nu^c$
 \bea
\lambda_e L \tilde \phi L^c + \lambda_N N H_R L^c \rightarrow  
\lambda_e  L H_u^* e^c  + \lambda_e L H_d^* \nu^c + \frac{\lambda_N}{\sqrt{2}} f N  \nu^c \ .
\label{eq:nu_masses}
\eea
These couplings leave three generations of exactly massless neutrinos which are primarily the usual left-handed partners of the Standard Model charged leptons. The massless neutrinos contain a small admixture of $N$ which alters their coupling to the $W$ and $Z$. The mixing angle is suppressed by $\lambda_e v/(\lambda_N \tan\beta f)$ which is sufficiently small ($\sim 10^{-4}$) even for the neutrino partner of the tau lepton. Acceptable Standard Model neutrino masses can be now be generated by adding a small violation of lepton number. For example, small majorana masses for $N$ induce small majorana masses for the Standard Model neutrinos through mixing.

\end{document}